\documentclass[aps,reprint,twocolumn,nofootinbib,prc,amsmath,amssymb]{revtex4-1}

\usepackage{graphicx}
\usepackage{marginnote}
\usepackage{float}
\usepackage{color}
\usepackage{scrextend}

\usepackage[latin9]{inputenc}
\usepackage[T1]{fontenc}
\usepackage[english]{babel}
\usepackage{graphicx}
\usepackage{subfigure}
\usepackage{cancel}
\usepackage{ulem}
\usepackage{amssymb}
\usepackage{epsfig}
\usepackage{appendix}
\usepackage{graphicx}
\usepackage{bm}
\usepackage{dcolumn}
\usepackage{multirow}
\usepackage{float}
\usepackage{listings}
\usepackage{color}
\usepackage{todonotes}

\include{srctex}

\newcommand{\ignore}[1]{}

\begin{document}

\title{Spontaneous twisting and shrinking of carbon nanotubes\vspace{0ex}}

\author{V\'i{}t Jakubsk\'y}
\email{jakub@ujf.cas.cz}
\author{Axel P\'erez-Obiol}
\email{perez-obiol@ujf.cas.cz}
\affiliation{Department of Theoretical Physics, Nuclear Physics Institute, 25068  \v Re\v z, Czech Republic}

\date{\today}

\begin{abstract}
Deformations of single-wall carbon nanotubes are investigated within the  tight-binding model with deformation-dependent hopping energies. We show that the nanotubes tend to twist and shrink spontaneously at zero temperature. The explicit values of the deformation parameters  are computed for a wide range of nanotubes with varying diameter and chirality.  The changes of the spectral gap associated with the spontaneous deformation are shown to depend on the chirality of the nanotubes.
\end{abstract}
\maketitle

\section{Introduction}

Carbon nanotubes are considered as one of the most promising materials for the future of electronic devices \cite{peng}. They can be both metallic or semiconducting, depending on the orientation of the lattice in their shells \cite{Blase},
and their electronic properties can be further altered through mechanical deformations \cite{KaneMele}.

Charge carriers in metallic nanotubes can propagate over long distances without being scattered by impurities \cite{andoI,White,tans,liang}. This effect is understood as a manifestation of Klein tunneling in the carbon nanostructures \cite{Katsnelson,Young}. However, when the gap opens, the electrons acquire an effective mass, the back-scattering on the impurities takes place, and the conductance decreases \cite{suzuura}. 

In the zone-folding approach \cite{saito}, the electronic properties of carbon nanotubes are deduced from those of planar graphene. Within this scheme, the hopping energies associated to the three nearest neighbor bonds are considered to have the same value. It was predicted that one third of the nanotubes are metallic, while the rest contain a gap in their spectrum and are semiconducting. In a more realistic description, the orientation of the bonds in the shell of the nanotube is taken into account \cite{KaneMele}, \cite{Mahan}. It was shown that, in this case, a gap proportional to the inverse of the nanotube radius, $\frac{1}{R}$, opens both in zig-zag and chiral metallic nanotubes, leaving the armchair nanotubes as the only candidates for a genuine one-dimensional metal. 

Long time ago, it was found by Peierls \cite{Peierls} that one-dimensional metals are unstable under distortions that open a gap at  temperatures $T<T_c$. This phenomenon was studied for metallic carbon nanotubes \cite{Mintmire}, \cite{Dumont}, \cite{Connetable}, \cite{Figge1}, \cite{Figge2}, and it was shown that the dimerization of the interatomic bonds (Kekul\'e distortion) opens a gap even in the armchair nanotubes.
Estimates of the transition temperature $T_c$, in which the deformation energy is too large to be compensated by the gap opening, vary from $T_c<1K$ in \cite{Mintmire} to several tens of kelvins in \cite{Chen} or even to the room temperature for thin nanotubes \cite{Connetable}. In \cite{Figge2} it was found that the transition temperature decays exponentially with the radius.
 Later, it was argued that electron-electron interactions \cite{Connetable} can enhance the Peierls distortion such that the transition temperature behaves rather like $\frac{1}{R^3}$.
The dimerization of a nanotube is associated with the optical phonons, in which the atoms of the two triangular sublattices oscillate with opposite phase.
It was discussed e.g. in \cite{Figge2} that the acoustic phonons, called twistons, can open a gap in the spectrum. 

In nature, axially twisted single-wall carbon nanotubes can be found in bundles of nanotubes \cite{Clauss}. They have also been realized in experiments, where they have served as torsional strings \cite{meyer}.
By altering the spectral gap, the twist affects the transport properties of the nanotubes. In particular, the twist can cause conductance oscillations, producing metal-semiconductor transitions \cite{JoselovichII}. The existence of topologically nontrivial static configurations ---solitwistons---, that minimize the free energy at $T=0$, were discussed in \cite{SSH}, \cite{TLM}, \cite{Figge1}, \cite{Figge2}.

In the current article, we address the question of whether single-wall carbon nanotubes, both metallic and semiconducting, get spontaneously twisted and dilated at zero temperature. We employ a generalization of the zone-folding approach, where the tight-binding model assumes a modification of the hopping energies between the nearest neighbor atoms due to the curvature, chirality and elastic deformation (the twist and the dilation) of the nanotube \cite{wang}. 
The article is organized as follows: in the next section, the model of the axially twisted and dilated nanotube, based on the tight-binding Hamiltonian, is presented. In section \ref{S2}, we study the case of infinitely long nanotubes, which prove to be a relevant approximation of the realistic systems. The twists, dilations, and energy gaps corresponding to the most stable configurations are found numerically for different radiuses and chiralities. The results are discussed in the section \ref{S5}.  

\section{Tight-binding Hamiltonian\label{S2}}
The crystal structure of an undeformed nanotube $(n,m)$ is defined by its circumference (wrapping) vector $\mathbf{C^0_h}=n \mathbf{a}_1+m \mathbf{a}_2$, where $\mathbf{a}_1$ and $\mathbf{a}_2$ are the primitive translation vectors of the hexagonal lattice. 
The associated translation vector $\mathbf{T^0_h}=t_1\mathbf{a_1}+t_2\mathbf{a_2}$ along the nanotube is fixed as the shortest vector that satisfies $\mathbf{C^0_h} .\mathbf{T^0_h}=0$ with integer valued $t_1$ and $t_2$. It can be written explicitly as $\mathbf{T^0_h}=\frac{2m+n}{d}\mathbf{a_1}-\frac{2n+m}{d}\mathbf{a_2}$ where $d$ is the greatest common divisor of $2m+n$ and $2n+m$, see \cite{saito} for more details.

The surface of the nanotube is parametrized as $\mathbf{\tau^0}
=(\xi^0,\zeta^0)=(R\Theta^0,\zeta^0)$, where $\xi^0\in[0,|\mathbf{C^0_h}|]$, $\Theta^0\in[0,2\pi]$ and $\zeta^0\in [0,L]$, $L$ being the length of the nanotube. In these coordinates, the circumference and the translation vectors 
are $\mathbf{C^0_h}=(|\mathbf{C^0_h}|,0)^t$ and $\mathbf{T^0_h}=(0,|\mathbf{T^0_h}|)^t$, and the associated reciprocal vectors are
$\mathbf{K^0_1}=\left(\frac{2\pi}{|\mathbf{C^0_h}|},0\right)$ and $\mathbf{K^0_2}=\left(0,\frac{2\pi}{|\mathbf{T^0_h}|}\right)$.

The chiral angle $\chi$ between the nearest neighbor vector and the circumference vector $\mathbf{C_h}$ reads $\chi=\arcsin\frac{n-m}{2\sqrt{n^2+m^2+nm}}.$ Due to the symmetry of the hexagonal lattice, we  can take $\chi\in [0,\pi/6]$.
The nearest neighbor vectors $\mathbf{\tau^0_i}=(\xi^0_i,\zeta^0_i)$, $i=1,2,3$, are written as
\begin{align}
\label{coordinates}
\xi^0_1=&-a_{cc}\cos\chi,
& \zeta^{0}_1=&-a_{cc}\sin\chi,
\\\nonumber
\xi^0_2=&a_{cc}\cos\left(\frac{\pi}{3}-\chi\right),
&\zeta^0_2=&-a_{cc}\sin\left(\frac{\pi}{3}-\chi\right),
\\\nonumber
\xi^0_3=&a_{cc}\cos\left(\frac{\pi}{3}+\chi\right),
& \zeta^0_3=&a_{cc}\sin\left(\frac{\pi}{3}+\chi\right),
\end{align}
where $a_{cc}= \mathrm{0.142}$~nm is the distance between the carbon atoms. 

Dilation and axial twisting of the nanotube change the coordinates from $\mathbf{\tau^0}$ to $\mathbf{\tau}$ through the matrix $\mathbf{D}$,
\begin{equation}\label{deftr}
 \mathbf{\tau}=\mathbf{D}(\delta,\epsilon)\mathbf{\tau^0},\quad \mathbf{D}(\delta,\epsilon)=\left(\begin{array}{cc}1-\nu \epsilon & R_{\epsilon}\delta\\ 0& 1+\epsilon\end{array}\right),
\end{equation}
where  $\nu=0.165$ is the Poisson ratio for graphite \cite{Blakslee}, $\delta$ is the twist per length in units of $\mbox{rad}/\mbox{nm}$, and $\epsilon$ is the dilation parameter. The radius of the dilated nanotube is $R_{\epsilon}=(1-\nu\epsilon)R$, where $R=\frac{\sqrt{3}}{2\pi}\sqrt{n^2+m^2+nm}$. We assume that $R\delta<<1$ and $\epsilon<<1$.
A deformation changes both the circumference vector and the translation vector of the nanotube,
\begin{equation}
 \mathbf{C_h}=\mathbf{D}(\delta,\epsilon)\mathbf{C_h^0},\quad \mathbf{T_h}=\mathbf{D}(\delta,\epsilon)\mathbf{T_h^0},
\end{equation}
as well as the reciprocal vectors,
\begin{equation}
 \mathbf{K_1}=\mathbf{K^0_1}\,\mathbf{D}(\delta,\epsilon)^{-1},\quad \mathbf{K_2}=\mathbf{K^0_2}\,\mathbf{D}(\delta,\epsilon)^{-1}. 
\end{equation}

The free electrons in the nanotube are labeled by three quantum numbers: the longitudinal momentum $k$, the angular quantum number $\alpha$ that acquires integer values and fixes the angular momentum $k_{\xi}=\frac{\alpha}{R_{\epsilon}}$, and the spin of electrons $\sigma=\pm\frac{1}{2}$. 

In principle, the longitudinal momentum also gets quantized  by the boundary conditions that depend on the particular ending of the nanotube \cite{Brey,AkhBee}.
However, in our work, we assume that the nanotube is infinite, and that $k$ takes a continuum of values. This is a reasonable approximation since very long nanotubes can be isolated in experiments \cite{Zheng}, \cite{Wang}.

The tight-binding Hamiltonian can be written as
\begin{align}
 H_{tb}=&-\sum_{k,\alpha,\sigma}(a_{k,\alpha,\sigma}^\dagger,b_{k,\alpha,\sigma}^{\dagger})
 \\&\times\nonumber
 \left(\begin{array}{cc}0&\sum_{\eta=1}^3t_\eta e^{i\varphi_\eta}\\\sum_{\eta=1}^3t_{\eta}e^{-i\varphi_\eta}&0\end{array}\right)\left(\begin{array}{c}a_{k,\alpha,\sigma}\\b_{k,\alpha,\sigma}\end{array}\right).
\end{align}
Here, the operators $a_{k,\alpha,\sigma}^{\dagger}$ and $b_{k,\alpha,\sigma}^{\dagger}$ create electrons with the longitudinal momentum $k$, angular number $\alpha$, and spin $\sigma$ on the triangular  sublattices $A$ and $B$, respectively. The phase factors $\varphi_{\eta}$ depend on the coordinates of the nearest neighbor vectors $\tau_{\eta}$. Defining $\mathbf{k}=(k_\xi,k)$, they read \cite{Mahan}, \cite{wang},
\begin{equation}\label{varphi}
 \varphi_{\eta}=\mathbf{k}\cdot\mathbf{\tau_\eta}=\alpha\Theta_{\eta}+k \zeta_{\eta}, \quad \Theta_{\eta}=R_{\epsilon}^{-1}\xi_{\eta}.
\end{equation}
The hopping energies $t_\eta$ are altered due to the curvature of the cylindrical surface and due to the mechanical deformation. They can be expressed in terms of the metric and curvature tensors $g_{ij}$ and  $K_{ij}$,

\begin{equation}\label{g}
\mathbf{g}=\left(\begin{array}{cc}(1-\nu\epsilon)^2&(1-\nu\epsilon)R_\epsilon\delta\\(1-\nu\epsilon)R_\epsilon\delta&(1+\epsilon)^2+R_\epsilon^2\delta^2\end{array}\right),~ \mathbf{K}=\left(\begin{array}{cc}\frac{1}{R_\epsilon}&0\\0&0\end{array}\right),
\end{equation}
as
\begin{align}\label{teta}
t_\eta=&\gamma_0\left(1-\frac{1}{8}\tau^{0}_{\eta i}\tau^{0}_{\eta j}K_{jk}K_{il}g^{lk}\right)
-\frac{\kappa\beta\gamma_0}{2a_{cc}^2}\tau_{\eta i}^0\tau_{\eta j}^0(g_{ij}-\delta_{ij})
\nonumber\\
=&\gamma_0\left(1-\frac{(1-\nu\epsilon)^2}{8R_{\epsilon}^2}(\xi_\eta^0)^2\right)
-\frac{\kappa \beta \gamma_0}{2a_{cc}^2}\left[
(\xi_{\eta}^0)^2\nu\epsilon(\nu\epsilon-2)
\nonumber\right.\\&\left.
+2\xi_{\eta}^0\zeta_\eta^0R_{\epsilon}\delta(1-\nu\epsilon)
+(\zeta_\eta^0)^2(R^2_{\epsilon}\delta^2+2\epsilon+\epsilon^2)\right].
\end{align}
$\gamma_0=2.7$ eV is the nearest-neighbor hopping parameter for planar graphene
\cite{reich}.
The constant
$\beta=-\frac{\partial \ln \gamma_0}{\partial \ln a_{cc}}$ expresses dependence of the nearest-neighbor hopping parameter on the interatomic distance. Its value is estimated to be $\beta\sim 2$ \cite{andoII}. However, there are experiments that indicate that its value can be smaller \cite{Hertel}. 
The coefficient $\kappa=\frac{\mu}{\sqrt{2}(\lambda+\mu)}$, with $\lambda=240$ $\mathrm{eV/nm^2}$ and $\mu=990$ $\mathrm{eV/nm^2}$ being the graphene Lam\'e parameters \cite{zakharchenko}, \cite{vozmediano}. 
The correction term in the first bracket  in (\ref{teta}) is due to the curvature of the nanotube, while the second term arises due to the deformation, see \cite{wang} for more details. 

Let us define $b'_{k,\alpha,\sigma}=b_{k,\alpha,\sigma}\frac{\sum_{\eta}t_{\eta}e^{-i\varphi_\eta}}{|\sum_{\eta}t_{\eta}e^{-i\varphi_\eta}|}$. Then $H_{tb}$ can be diagonalized in terms of the operators $u_{k,\alpha,\sigma}$ and $v_{k,\alpha,\sigma}$, \cite{Mahan}, \cite{wang},
\begin{align}
\label{Htb2}
& H_{tb}=\sum_{k,\alpha}\left(E_{k\alpha}^+u_{k,\alpha,\sigma}^{\dagger}u_{k,\alpha,\sigma}+E_{k,\alpha}^-v_{k,\alpha,\sigma}^\dagger v_{k,\alpha,\sigma}\right),
\nonumber \\\nonumber
&u_{k,\alpha,\sigma}=\frac{1}{\sqrt{2}}(a_{k,\alpha,\sigma}+b'_{k,\alpha,\sigma}),
\\ 
&v_{k,\alpha,\sigma}=\frac{1}{\sqrt{2}}(a_{k,\alpha,\sigma}-b'_{k,\alpha,\sigma}),
\end{align}
where $E_{k\alpha}^{\pm}$ is defined as
\begin{align}
E^{\pm}(k,\alpha)=&
\pm\left[t_1^2+t_2^2+t_3^2+2t_1t_2\cos(\varphi_2-\varphi_1)
\right.\\&\nonumber\left.
+2t_2t_3\cos(\varphi_3-\varphi_2)+2t_1t_3\cos(\varphi_3-\varphi_1)\right]^{\frac12}.
\end{align}
Both $t_{\eta}$ and $\varphi_{\eta}$ depend on the deformation parameters $\delta$ and $\epsilon$, while only $\varphi_{\eta}$ depends on
 $k$ and $\alpha$.

\begin{figure}[t]
	\centering
	\includegraphics[scale=0.2]{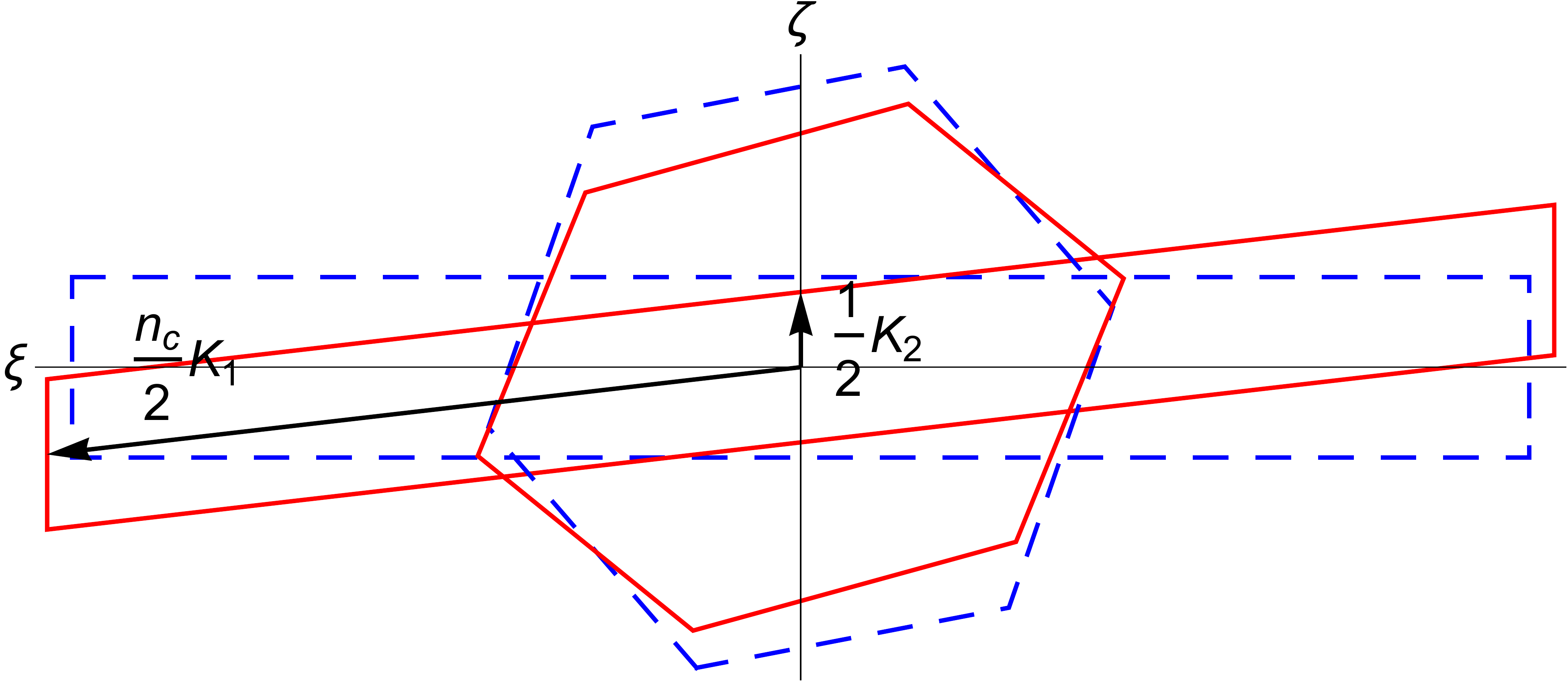}
	\caption{Primitive rectangular and hexagonal cells in the reciprocal space
	for a (16,4) nanotube. The blue and dashed cells correspond to the non-deformed
	case, while the solid and red cells represent a deformation with $\epsilon>0$ and $\delta>0$.
	A finite dilation $\epsilon>0$ prolongs and shrinks the cells in the $\xi$ and $\zeta$
	directions, respectively, while a finite twist $\delta>0$ shears the cells in the $\xi$
	direction.}
	\label{fig:cells}
\end{figure}

In the zone-folding of the dispersion relations \cite{saito}, the momentum $\mathbf{k}$ acquires the form
\begin{align}
 \mathbf{k}=&
 \alpha\mathbf{K_1}+s\mathbf{K_2},
\end{align}
with
\begin{align*}
 \alpha\in\left\{-\frac{n_c}{2}+1,\dots,\frac{n_c}{2}\right\},~
  s\in[-1/2,1/2],
\end{align*}
and where $n_c=t_1m-t_2n$. Note that in a deformed nanotube, $\mathbf{K_1}$ and $\mathbf{K_2}$ are not perpendicular. The twist and deformation cause tilting of $\mathbf{K_1}$ and shrinking of $\mathbf{K_2}$, see Fig.~\ref{fig:cells}. The longitudinal momentum can be obtained as $k=\mathbf{k}\cdot\frac{\mathbf{K_2}}{|\mathbf{K_2}|}$.

\section{Spontaneous deformations at $T=0$\label{2}}

The Hamiltonian for a carbon nanotube can be divided into
the tight-binding Hamiltonian, $H_{tb}$,
 and the Hamiltonian of the crystal lattice,
$H_{lat}$,
\begin{equation}
 H(\delta,\epsilon)=H_{tb}+H_{lat}.
\end{equation}
$H_{tb}$ describes the $\pi$-electrons of the nanotube, as in Eq.~(\ref{Htb2}),
while $H_{lat}$ characterizes the atomic displacements and vibrations ---the phonons---.
The interaction of $\pi$-electrons with phonons is rather implicit in $H_{tb}$.
It can be traced in the form of the hopping energies $t_{\eta}$ and the phase factors $\varphi_{\eta}$, see Eqs.~(\ref{varphi}) and  (\ref{teta}) respectively.
The lattice Hamiltonian can be further divided into 
\begin{equation}
H_{lat}=U(\delta,\epsilon)+H_{vib}.
\end{equation}
$U(\delta,\epsilon)$ represents the energy stored in the static deformation, and $H_{vib}$ the energy of the vibrations of the atoms around their equilibrium positions.

The elastic energy depends on how the atoms in the lattice are shifted due to deformation,
which is indicated by the deformation vector of every atom, defined as
 $\mathbf{u}(\xi,\zeta)=\mathbf{\tau}-\mathbf{\tau_0}=(-\nu\epsilon \xi+R_\epsilon \delta\zeta, \epsilon\zeta)$, see Eq.~(\ref{deftr}). 
In the continuum limit, $U(\delta,\epsilon)$ can be expressed in terms of the strain tensor,
which describes how $\mathbf{u}(\xi,\zeta)$ changes through the nanotube,
\begin{equation}
u_{ij}=\frac{1}{2}\left(\partial_i u_j+\partial_j u_i\right)=\left(\begin{array}{cc}-\nu\epsilon&\frac{1}{2}R_{\epsilon}\delta\\\frac{1}{2}R_{\epsilon}\delta&\epsilon\end{array}\right).
\end{equation} 
The static energy $U(\delta,\epsilon)$ as a function of the strain tensor is then
\begin{align}
U(\delta,\epsilon)=&\int dxdy \frac{1}{2}\left[(\lambda+\mu)(u_{xx}+u_{yy})^2
\right.\\\nonumber &\left.
+\mu \left((u_{xx}-u_{yy})^2+4u_{xy}^2\right)\right]
\\\nonumber
=& \pi\,R\, L\left[
\mu\,R_{\epsilon}^2\,\delta^2+
\lambda (1-\nu)^2\epsilon^2+2\mu(1+\nu^2)\epsilon^2\right].
\end{align}

The equilibrium configuration of a particular nanotube
corresponds to the absolute minimum in its free energy.
Fixing the chemical potential of the electrons and phonons to zero \cite{fang}, the free energy ${F}$ in terms of the partition function reads,
\begin{equation}\label{ff}
F(\delta,\epsilon)=-k_B T\log Z(\delta,\epsilon),\quad Z(\delta,\epsilon)=\mbox{Tr} e^{-\frac{ H(\delta,\epsilon)}{k_B T}},
\end{equation}
where $k_B$ is the Boltzmann constant.
For $T=0$, it coincides with the
inner energy 
$E_{in}$,
\begin{align}\label{FF}
F=&E_{in}=U+E_{vib}
\\ & \nonumber
-\frac{L_{\epsilon}}{\pi}\sum_{\alpha=-\frac{n_c}{2}+1}^{\frac{n_c}{2}}\int_{c(\alpha)}^{c(\alpha)+|K_2|} E^+(k,\alpha)dk,
\end{align}
where 
$c(\alpha)=-\frac{|K_2|}{2}-\frac{\alpha\,\delta}{(1+\epsilon)}$
and $L_{\epsilon}=(1+\epsilon)L$ is the physical length.
  The first term represents the energy of the static deformation. The third term corresponds to the energy of the Valence band with all its levels occupied and multiplied by 2 to account for
  spin degeneracy. The second term, $E_{vib}$, is the sum of all phonon energies, which at $T=0$ corresponds to the zero-point vibrations. 
In the harmonic approximation, the phonon frequencies are functions of phonon momenta but are independent of the temperature and both $\delta$ and $\epsilon$.
 
The condition for the minimum of $F$ can be written as
\begin{align}
\label{par}
 \partial_\delta F(\delta,\epsilon)=\partial_\epsilon F(\delta,\epsilon)=0,\quad \partial_{\delta\delta}^2F(\delta,\epsilon)>0,\quad 
 \\\nonumber
\partial^2_{\delta\delta}F(\delta,\epsilon)\partial^2_{\epsilon\epsilon}F(\delta,\epsilon)-(\partial^2_{\delta\epsilon}F(\delta,\epsilon))^2>0.
\end{align}
It implies that the terms independent of either $\delta$ or $\epsilon$ do not alter the position of the extremum of $F$ in the  $(\delta,\epsilon)$ plane.  

\begin{figure}[t]
\centering
\includegraphics[width=\columnwidth]{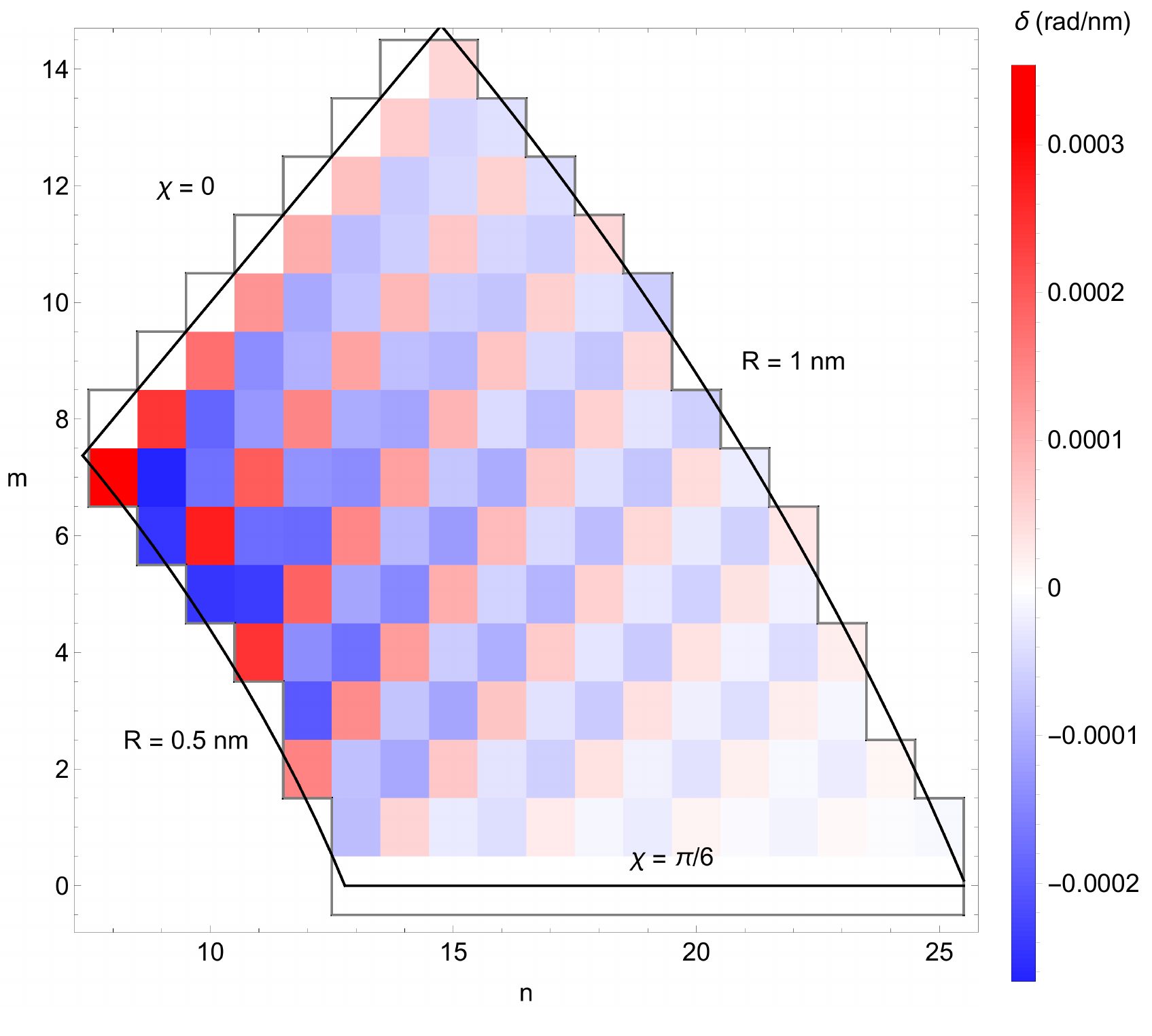}
\caption{(Color online) Spontaneous twist for all nanotubes with 0.5 nm < R < 1 nm
organized by their chiral indices (n,m).}
\label{fig:twists}
\end{figure}

Numerically minimizing $F/L$ with respect to $\delta$ and $\epsilon$
 we find the most stable configuration for a particular carbon nanotube (n,m).
 We have considered all the possible nanotubes with radiuses in the range 0.5 nm < R < 1 nm.
 Radiuses smaller than 0.5 nm are outside the scope of the model we use,
which is only valid for small curvatures \cite{LammertCrespi}.
Nanotubes with radiuses larger than 1 nm seem to collapse \cite{benedict,xiaocollapse}.
 In Fig.~\ref{fig:twists}, the magnitude of the twists is shown depending on the
 chiral indices (n,m). All but zigzag and armchair nanotubes are found to spontaneously twist
 a few rad every 10 $\mathrm{\mu m}$.
 The nanotubes with $\mod(n-m,3)=1$
  twist in opposite directions than the rest
 ---note that nanotubes with fixed $\mod(n-m,3)$ are represented in Fig.~\ref{fig:twists} by the diagonals with slope 1---.
 The twists increase with decreasing chirality and radius, being the most twisted nanotubes those with
 the smallest, non-zero chiral angle and radius.
 The fact that zigzag and armchair nanotubes do not twist can be understood by expanding
 the free energy in powers of $\delta$ up to $\delta^2$.
 On the one hand, these nanotubes are axially symmetric and the linear terms in $\delta$ always vanish.
 On the other hand, the quadratic  term is found numerically to be always positive, obtaining thus a minimum at $\delta=0$, independently of the radius and the shrinking.
 
\begin{figure}[t]
\centering
\includegraphics[width=\columnwidth]{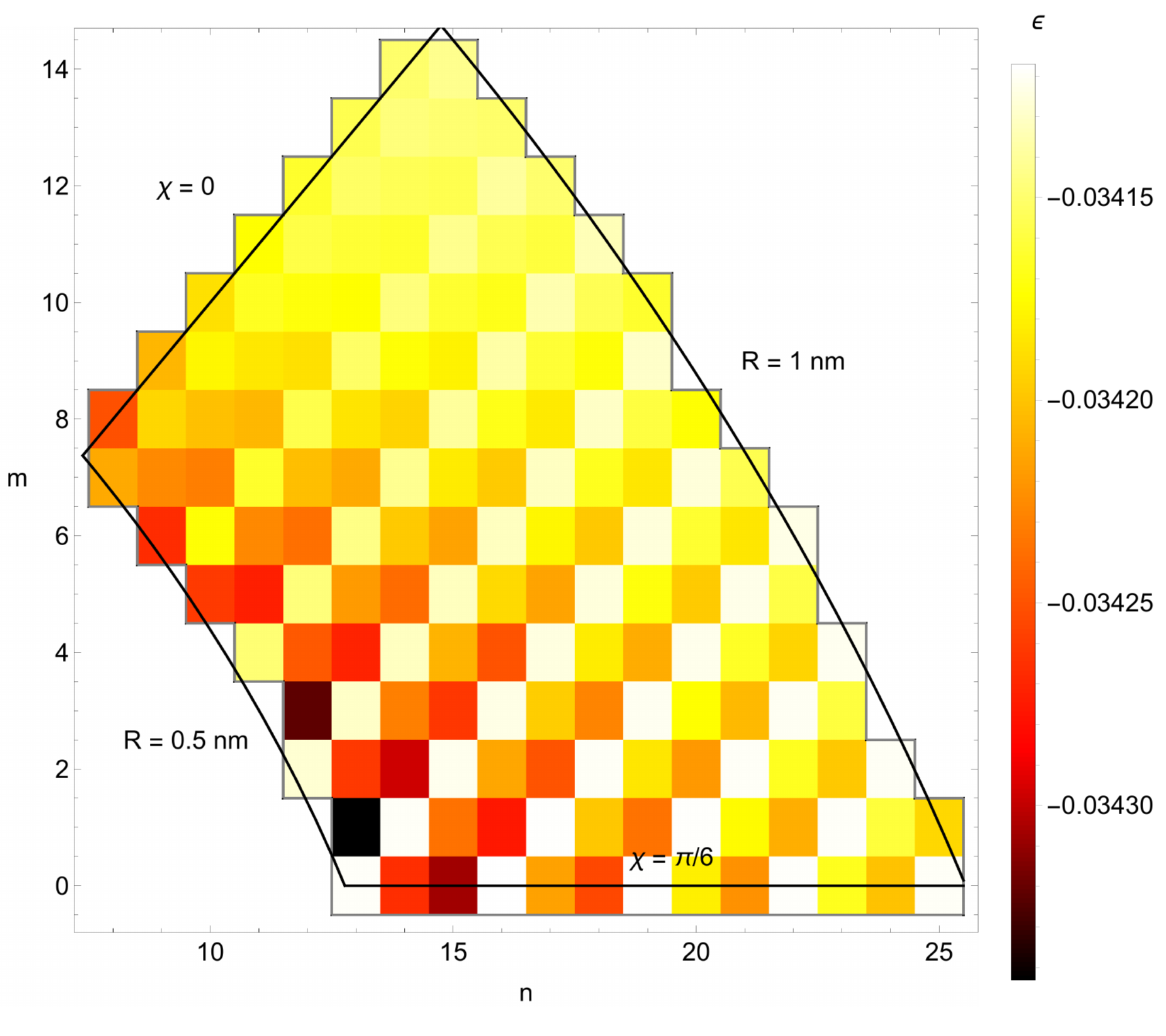}
\caption{(Color online) Spontaneous shrink for all nanotubes with 0.5 nm < R < 1 nm
organized by their chiral indices (n,m).}
\label{fig:shrinks}
\end{figure}

In Fig.~\ref{fig:shrinks} the magnitude of the parameter $\epsilon$
is plotted for all the nanotubes considered. In this case,
we find that all nanotubes shrink about $3.4\%$.
Within the small differences, the shrinkages
decrease with the radius,
their specific magnitude depending mainly on the value of $\mod(n-m,3)$.
Nanotubes with $\mod(n-m,3)=0,2$ increase
their shrinkage with chirality,
while the opposite happens with nanotubes with $\mod(n-m,3)=1$.


Once the most stable configurations for the set of nanotubes with 
0.5 nm < R < 1 nm are found, it is natural to ask how their spontaneous
deformations affect their dispersion relation, and in particular their
energy gaps.
The couplings $t_\eta$ appearing in the dispersion relation are
modified differently depending on the chirality and the deformation
of the nanotube.
More precisely, they depend on the coordinates of the corresponding
coupling vectors $\tau_\eta$ (Eq.~\ref{coordinates}), which define the separation and direction of the pair of nearest neighbors coupled by $t_\eta$.
Neglecting the twist and terms of order ${\cal O}(\epsilon^2)$
or superior, the difference between the couplings with and without
deformation is
\begin{align}
	t_\eta^\epsilon-t_\eta^0=&
	\frac{\kappa\beta\gamma_0}{a_{cc}^2}\epsilon
	\left(\nu(\xi_\eta^0)^2-(\zeta_\eta^0)^2\right)+{\cal O}(\epsilon^2).
\end{align}
Since $\epsilon\simeq -0.034$ for all nanotubes, this equation
shows that, with spontaneous shrinking,
the coupling between two first neighbors becomes larger
the more they are oriented in the axial direction.
The quantity $t_1^2+t_2^2+t_3^2$ does not depend
on the chiral angle, and is affected mainly by $\epsilon$.
We find that it is about 3\% larger than
in the case without deformation (corresponding approximately to $3\gamma_0^2$).

\begin{figure}[t]
\centering
\includegraphics[width=\columnwidth]{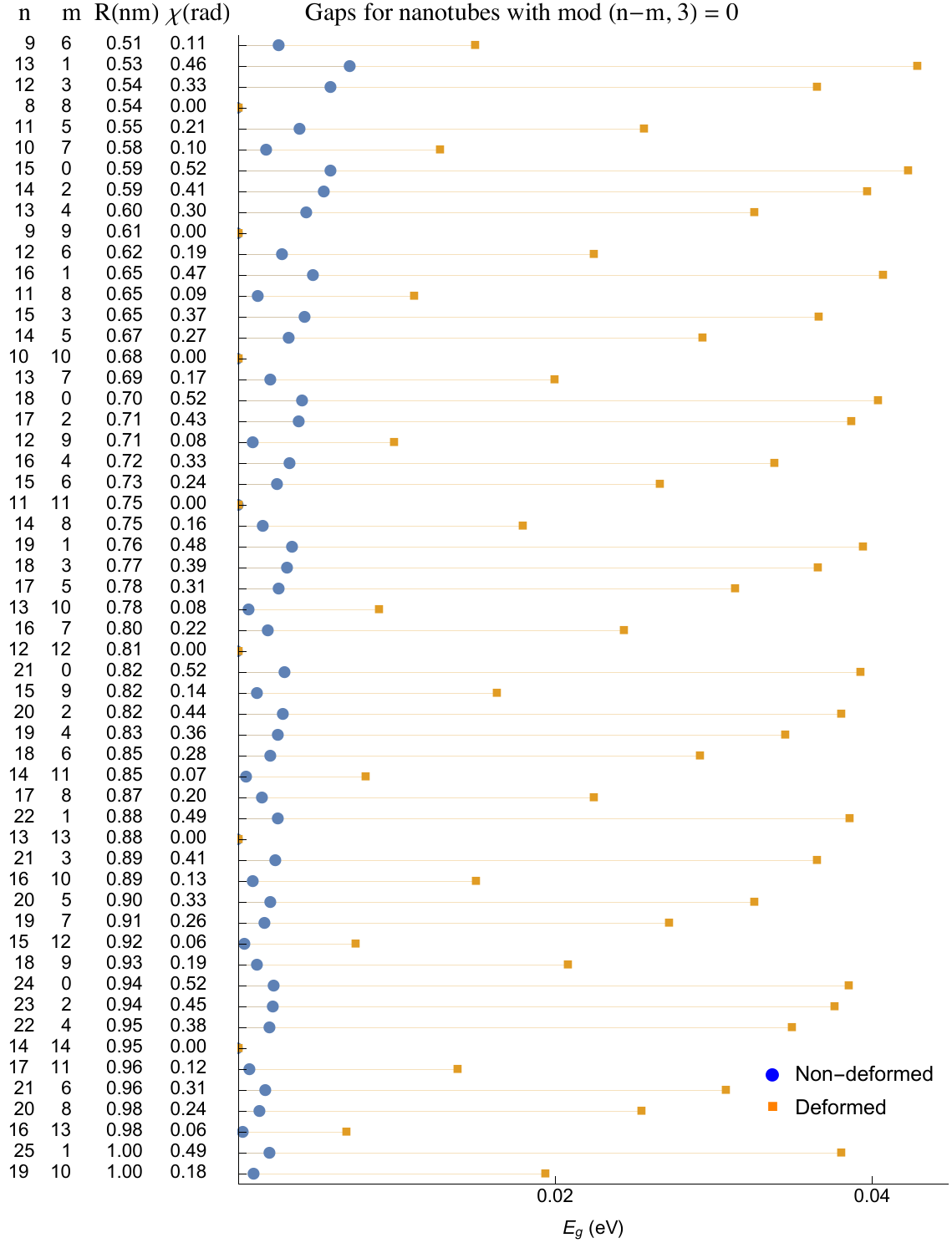}
\caption{(Color online) Gap for all nanotubes with 0.5 nm < R < 1 nm,
chiral indices such that $\mod(n-m,3)=0$,
and ordered by increasing radius.
The  circles and  squares correspond, respectively,
 to the gaps without and with deformation}
\label{fig:gaps0}
\end{figure}

\begin{figure}[ht]
\centering
\includegraphics[width=\columnwidth]{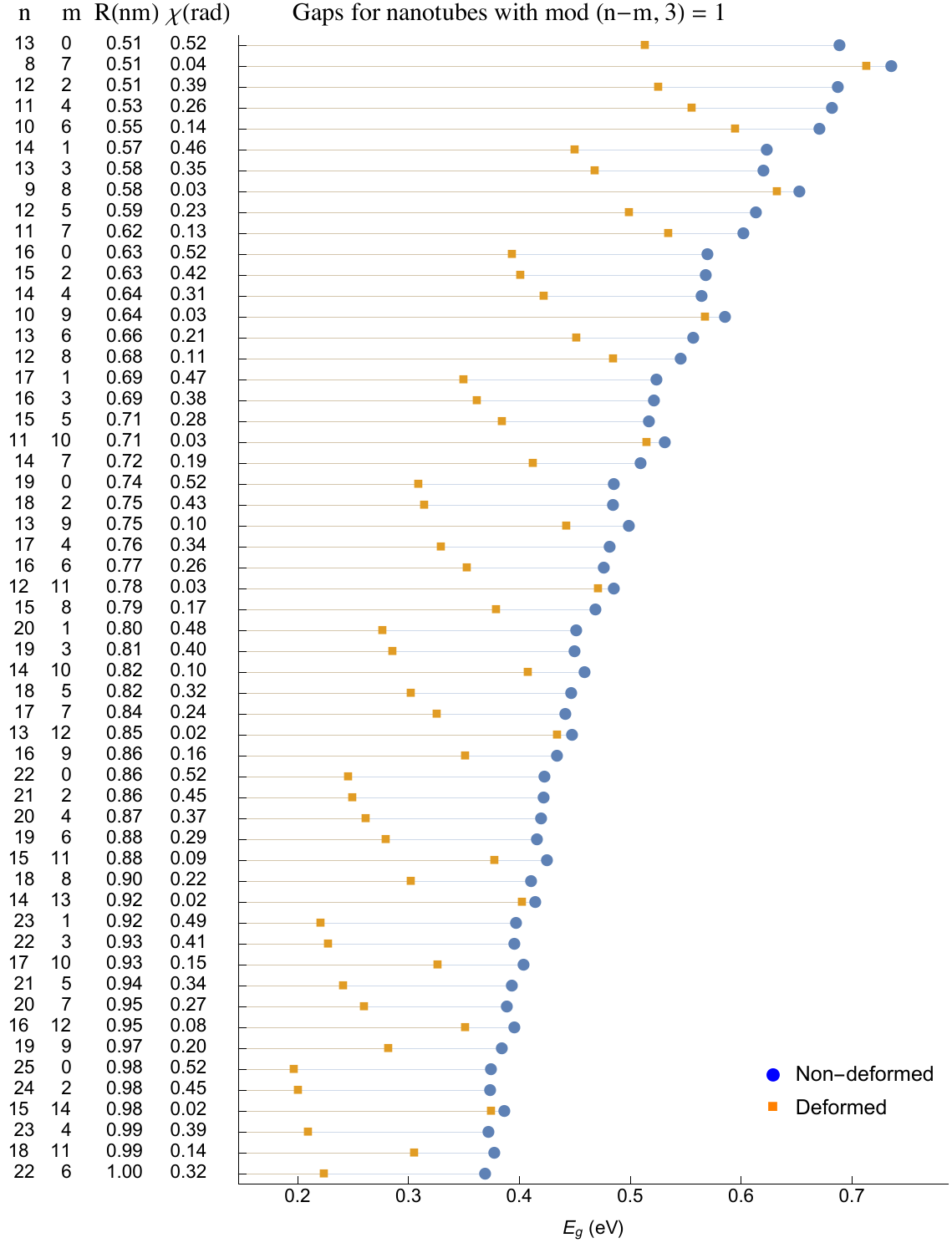}
\caption{(Color online) Gap for all nanotubes with 0.5 nm < R < 1 nm,
chiral indices such that $\mod(n-m,3)=1$,
and ordered by increasing radius.
The circles and squares correspond, respectively,
 to the gaps without and with deformation}
\label{fig:gaps1}
\end{figure}

\begin{figure}[t]
\centering
\includegraphics[width=\columnwidth]{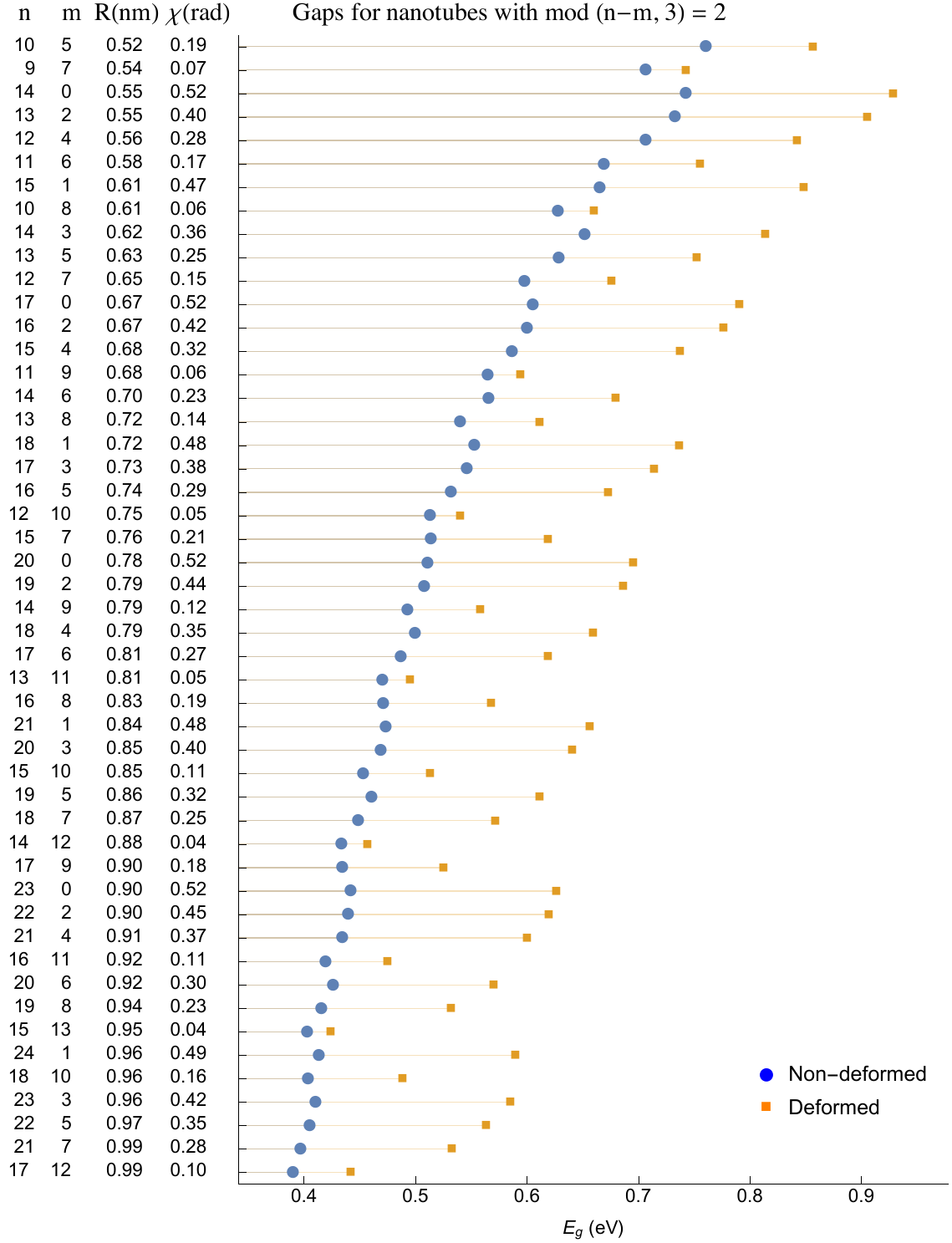}
\caption{(Color online) Gap for all nanotubes with 0.5 nm < R < 1 nm,
chiral indices such that $\mod(n-m,3)=2$,
and ordered by increasing radius.
The circles and squares correspond, respectively,
 to the gaps without and with deformation}
\label{fig:gaps2}
\end{figure}

The energy gaps are computed by finding the minima of $E^+(k,\alpha)$ and multiplying their values by two.
We compute them for all the nanotubes considered,
first fixing $\delta$ and $\epsilon$ to zero and second
to the values corresponding to the spontaneous deformation.
In both cases the magnitude of the gaps strongly depends on
the value of $\mod(n-m,3)$, which determines how far
the momentum lines are separated from the Dirac point.
In the case of non-deformed nanotubes, the shortest distance between
the Dirac point and the nearest momentum line is
0 for $\mod(n-m,3)=0$, and $\frac{a_{cc}\gamma_0}{3R_\epsilon}$ for $\mod(n-m,3)=1,2$. In Figs.~\ref{fig:gaps0}, \ref{fig:gaps1}, and \ref{fig:gaps2} we plot the gap values
for the nanotubes with, respectively, $\mod(n-m,3)=0,1,$ and 2.

Nanotubes with $\mod(n-m,3)=0$ have the smallest gap,
of the order of meV in the non-deformed case.
When considering the spontaneous deformation
for this group of nanotubes, the gaps grow by more than a factor $\sim10$ (except the armchair nanotubes).
The gap for the armchair nanotubes is exactly zero for both the non-deformed
and deformed cases. In this case the deformation does not modify the gap
because, as commented before, these nanotubes do not twist,
and the shrinking in armchair nanotubes only causes a displacement of the
the Dirac points  in the direction parallel to $\mathbf{K_2}$.

The gaps for the (undeformed) nanotubes with $\mod(n-m,3)=1,2$ are of the order of $\sim0.5$ eV, and their specific value  slightly decreases with the radius.
When spontaneously deformed, nanotubes with $\mod(n-m,3)=1$ decrease their
gap, whereas the ones with $\mod(n-m,3)=2$ increase it.
The particular magnitude of the change in the gap is more correlated
to the chirality than to the radius, as Figs. \ref{fig:gaps0}, \ref{fig:gaps1} and \ref{fig:gaps2} suggest ---the larger the chirality
the larger the increase/decrease in the gap---.


\section{Conclusions and outlook\label{S5}}

In our study we have considered how the free energy of the nanotube, consisting at zero temperature of the static elastic energy
and the energy of the electronic valence band, depend on the
twist and dilation of the nanotube. We found that at the equilibrium marked by the minimum of the free energy, all the considered nanotubes are spontaneously deformed.
At zero temperature, all nanotubes are shrunk
about $3.4\%$,
and all but the armchair and zigzag nanotubes are twisted,
even if only with a twist of the order of $\delta\sim 10^{-4}$ rad/nm.
Consequently, our model predicts a modification of the
energy gaps and hence the electronic properties of the nanotubes.
We found that the sign of the change of the gap depends on the value of $\mod(n-m,3)$ and the magnitude of the change strongly depends on the chirality of the nanotubes, see the Figs. \ref{fig:gaps0}, \ref{fig:gaps1} and \ref{fig:gaps2}.
This effect is especially relevant for the nanotubes where
$\mod(n-m,3)=0$, in which the gap grows considerably.

We considered deformations where the two triangular sublattices were shifted in the same directions. Dimerization (Kekul\'e distortion), where sublattices shift in opposite direction, were not included into our model. Hence, armchair nanotubes remain perfectly metallic,
and the rest are found to contain a gap
of the order of a few hundreds of meV.

The equilibrium configurations for the considered carbon nanotubes 
have been obtained through a minimization of their free energies,
which at $T=0$ and within the harmonic approximation,
do not depend on the lattice vibrations.
At finite temperatures, the excited electrons and phonons
also contribute to the free energy.
The analysis of spontaneous deformation at non-zero temperatures
requires a quantitative calculation of these contributions which goes beyond the scope of this article.

However, let us make a qualitative assessment on the relevant
effects as the temperature increases from $T=0$.
Considering $L$ to be finite again to avoid integrals in the formula,
the free energy at finite temperatures is
\begin{align}
\label{F}
F=&U-2\sum_{k,\alpha}E(k,\alpha)+\frac{1}{2}\sum_{q,p}
 \hbar \omega(q,p)
 \\&\nonumber
 -4k_BT\sum_k\log(1+e^{-E(k,\alpha)/(k_BT)})
 \\&\nonumber
 +k_BT\sum_{q,p}\log(1-e^{-\hbar\omega(q,p)/(k_BT)}).
\end{align}
$k$ and $q$ are the quantized longitudinal momenta of electrons and phonons,
and $\omega(q,p)$ are the phonon frequencies with momenta $q$ and $p$.
The last two terms do not contribute relevantly to the free energy at low temperatures due to the factor $k_BT$. In the harmonic approximation of lattice dynamics, the phonon frequencies are independent of the temperature and of the static deformations \cite{grosso}. Hence, within this framework, the spontaneous twisting and shrinking of the nanotube would persist despite increasing the temperature. Calculations beyond the harmonic approximation show that the independence of $\omega(q,p)$ on the deformation parameters is rather reasonable for very low temperatures, see Refs.~\cite{Figge1}, \cite{Figge2}, where the renormalized phonon frequencies were computed. It suggests that our model can be quite a sensible approximation at these temperatures.  When the temperature crosses a critical value, the frequencies depend on the deformation parameters. Then the third term in Eq.~(\ref{F}), corresponding to the zero-point vibrations, does not vanish when inserted in Eq.~(\ref{par}), and the spontaneous deformations become suppressed at larger temperatures.

\section*{acknowledgments}
This work was supported by the GA\v{C}R Grant No. 15-04301S and GA\v CR Grant No. 15-07674Y (Czech Republic).


\begin{thebibliography}{99}
\bibitem{peng} L.-M. Peng, Z. Zhang, and S. Wang, Materials Today {\bf17}, 433, (2014).

\bibitem{Blase} S. M.-M. Dubois, Z. Zanolli, X. Declerck, And J.-C. Charlier. Eur. Phys. J. B {\bf 72}, 1 (2009).

\bibitem{KaneMele} 
C. L. Kane, E. J. Mele, 
Phys. Rev. Lett. {\bf78} , 1932 (1997).

\bibitem{andoI} T. Ando and T. Nakanishi, J. Phys. Soc. Jap. {\bf 67}, 1704 (1998).

\bibitem{White} C.T. White and T. N. Todorov, Nature {\bf 393}, 240 (1998).

\bibitem{tans} s. J. Tans, M. H. Devoret, H. Dai, A. Thess, R. E. Smalley, L. J. Geerligs, and C. Dekker, Nature {\bf 386}, 474 (1997).

\bibitem{liang} W. Liang, M. Bockrath, D. Bozovic, J. H. Hafner, M. Tinkham, and H. Park, Nature {\bf 411}, 665 (2001).

\bibitem{Katsnelson} M. I. Katsnelson, K. S. Novoselov, and A. K. Geim, Nat. Phys. {\bf 2}, 620 (2006).

\bibitem{Young} A. F. Young and P. Kim, Nat. Phys. {\bf 5}, 222 (2009).

\bibitem{suzuura}H. Suzuura, Physica E: Low-dimensional Systems and Nanostructures {\bf 34}, 674 (2006).

\bibitem{saito}R. Saito, G. Dresselhaus, M. S. Dresselhaus, {\it Physical Properties of Carbon Nanotubes,} Imperial College Press (1998).

\bibitem{Mahan} 
G. D. Mahan, 
Phys. Rev. B  {\bf68}, 125409 (2003).

\bibitem{Peierls} R. E. Peierls, {\it Quantum mechanics of solids,} Oxford University Press (2001).

\bibitem{Mintmire} 
J. W. Mintmire, B. I. Dunlap, C. T. White, 
Phys. Rev. Lett.  {\bf68},  631 (1992).

\bibitem{Dumont} 
  G. Dumont, P. Boulanger, M. C\^ot\'e, M. Ernzerhof, 
Phys. Rev. B {\bf82}, 035419  (2010).

\bibitem{Connetable} 
D. Conn\'etable, G.-M. Rignanese, J.-C. Charlier, X. Blase, 
Phys. Rev. Lett. {\bf94}, 015503  (2005). 

\bibitem{Figge1} 
 M. T. Figge, M. Mostovoy, J. Knoester, 
Phys. Rev. Lett. {\bf 86}, 4572 (2001).



\bibitem{Figge2} 
 M. T. Figge, M. Mostovoy, J. Knoester, 
Phys. Rev. B {\bf65}, 125416 (2002).

\bibitem{Chen} 
Wei Chen, A. V. Andreev, A. M. Tsvelik, Dror Orgad, 
Phys. Rev. Lett. {\bf101}, 246802 (2008).

\bibitem{Clauss}W. Clauss, D. J. Bergeron, and A. T. Johnson, Phys. Rev. B {\bf 58}, 4266 (1998).

\bibitem{meyer}J. C. Meyer, M. Paillet, and S. Roth, Science {\bf309}, 1539 (2005).

\bibitem{JoselovichII} T. Cohen-Karni, L. Segev, O. Srur-Lave, S. R. Cohen, and E. Joselevich, Nat. Nanotechnol. {\bf 1}, 36 (2006).

\bibitem{SSH} W. P. Su, J. R. Schrieffer, A. J. Heeger, 
Phys. Rev. Lett.  {\bf42}, 1698 (1979).

\bibitem{TLM}
Hajime Takayama, Y. R. Lin-Liu, Kazumi Maki, 
Phys. Rev. B {\bf 21}, 2388 (1980). 

\bibitem{wang} 
Shaofeng Wang, Rui Wang, Xiaozhi Wu, Huili Zhang, Ruiping Liu, 
Physica E {\bf42}, 2250 (2010).



\bibitem{Blakslee}
O. L. Blakslee et al., J. Appl. Phys. {\bf41}, 3373 (1970);
E. J. Seldin and C. W. Nezbeda, J. Appl. Phys. {\bf41}, 3389 (1970).


\bibitem{Brey}L. Brey, H. A. Fertig,
Phys. Rev. B {\bf73}, 235411 (2006).

\bibitem{AkhBee}A. R. Akhmerov, C. W. J. Beenakker, 
Phys. Rev. B {\bf77}, 085423, (2008).


\bibitem{Zheng}
L. X. Zheng et al, 
Nat. Mater. {\bf3}, 673 (2004).

\bibitem{Wang}
Xueshen Wang et al,
Nano. Lett. {\bf9}, 3137 (2009).


\bibitem{reich} S. Reich, J. Maultzsch, C. Thomsen, and P. Ordejon,
Phys. Rev. B {\bf66}, 035412 (2002).

\bibitem{andoII}H. Suzuura, T. Ando, 
Phys. Rev B {\bf65}, 235412 (2002).

\bibitem{Hertel}T. Hertel, G. Moos, 
Phys. Rev. Lett. {\bf84}, 5002 (2000).

\bibitem{zakharchenko} K. V. Zakharchenko, M. I. Katsnelson, A. Fasolino, 
Phys. Rev. Lett. {\bf 102}, 046808 (2009).

\bibitem{vozmediano} M. A. H. Vozmediano, M. I. Katsnelson, F. Guinea, 
Phys. Rep. {\bf496}, 109 (2010).


\bibitem{fang} T. Fang, A. Konar, H. Xing, and D. Jena, Appl. Phys. Lett. {\bf91}, 092109 (2007).

\bibitem{LammertCrespi} 
P. E. Lammert, V. H. Crespi, 
Phys. Rev. B {\bf 61}, 7308 (2000).



\bibitem{benedict}
L. X. Benedict et. al, 
Chem. Phys. Lett. {\bf286}, 490 (1998).

\bibitem{xiaocollapse}
J. Xiao et. al, 
Nanotech. {\bf 18}, 395703 (2007).























\bibitem{grosso} G. Grosso, G. Pastori Parravicini,{\it Solid state physics,} Academic Press (2005) 



\end{thebibliography}
\end{document}